# Evidence of topological Kondo insulating state in MoTe$_2$/WSe$_2$ moiré bilayers

Zhongdong Han[1*], Yiyu Xia[2*], Zhengchao Xia[2], Wenjin Zhao[3], Yichi Zhang[1], Kenji Watanabe[4], Takashi Taniguchi[4], Jie Shan[1,2,3,5**], Kin Fai Mak[1,2,3,5**]

[1]Laboratory of Atomic and Solid State Physics, Cornell University, Ithaca, NY, USA
[2]School of Applied and Engineering Physics, Cornell University, Ithaca, NY, USA
[3]Kavli Institute at Cornell for Nanoscale Science, Ithaca, NY, USA
[4]National Institute for Materials Science, Tsukuba, Japan
[5]Max Planck Institute for the Structure and Dynamics of Matter, Hamburg, Germany

*These authors contributed equally
* *Email: jie.shan@cornell.edu; kinfai.mak@cornell.edu

**Topological Kondo insulators (TKIs) are topologically protected insulating states induced not by single-particle band inversions, but by the Kondo interaction between itinerant electrons and a lattice of local magnetic moments. Although experiments have suggested the emergence of three-dimensional (3D) TKIs in the rare earth compound SmB$_6$, its two-dimensional (2D) counterpart has not been demonstrated to date. Here we report experimental evidence of a TKI in angle-aligned MoTe$_2$/WSe$_2$ moiré bilayers, which support a Kondo lattice with topologically nontrivial Kondo interactions. We prepare in a dual-gated device a triangular lattice Mott insulator in the MoTe$_2$ layer Kondo-coupled to a half-filled itinerant band in the WSe$_2$ layer. Combined transport and compressibility measurements show that the prepared state supports metallic transport at high temperatures and, at low temperatures, an insulating bulk with conducting helical edge states protected by spin-S$_z$ conservation. The presence of Kondo singlets is further evidenced by their breakdown at high magnetic fields. Such behaviors are in stark contrast to the simple metallic state when the Mott insulator in the MoTe$_2$ layer is depleted by gating. Our results open the door for exploring tunable topological Kondo physics in moiré materials.**

## Main

A TKI is a novel state of matter, in which strong electronic correlation meets nontrivial band topology[1-3]. Its physics can be captured by the Anderson lattice model[4-6], in which the conduction electrons (c-electrons) support a dispersive energy band, the localized electrons (f-electrons) support a flat band, and the two bands are coupled by a hybridization term ($V$). Starting with a fully filled f-band (i.e. filling factor $\nu_f = 2$), electrons are transferred to the c-band as the on-site Coulomb repulsion ($U$) between the f-electrons is turned on adiabatically[3] (Fig. 1a,b). If $V$ is topologically trivial, the system is expected to crossover from a band insulator in the small $U$ limit, where $\nu_f \approx 2$ and $\nu_c \approx 0$, to a Kondo insulator in the large $U$ limit (i.e. the Kondo lattice limit), where $\nu_f \approx \nu_c \approx 1$. ($\nu_c$ is the c-electron filling factor.) If $V$ is topologically nontrivial, however, a topological phase transition from a band insulator to a TKI happens instead. Although experimental evidence of 3D TKIs[7-13] has been reported in the rare earth compound SmB$_6$, 2D TKIs have remained elusive so far. The emergence of moiré materials[14-18] not only provides an opportunity to realize 2D TKIs, but also to study and control the state continuously.

In this study, we demonstrate experimental evidence of a 2D TKI in the moiré semiconductor: angle-aligned MoTe$_2$/WSe$_2$ bilayers[19-21] (Mo- and W- for short from now on). The 7% lattice mismatch between the transition metal dichalcogenides (TMD) layers creates a honeycomb moiré lattice with a period of about 5nm. Density functional theory calculations[22] have established that the A and B sublattice sites of the honeycomb are occupied by Wannier orbitals from the Mo- and W-layer, respectively, and the two orbitals are coupled by an interlayer hopping term $V$. The electronic states of interest come from the time-reversal conjugate K and K' valley states near the valence band maxima in each TMD layer, which are spin-split by a large Ising spin-orbit coupling[18, 23]. Due to the heavier hole mass and the stronger moiré potential in the Mo-layer, the topmost moiré valence band of the Mo-layer is substantially flatter than that of the W-layer[22, 24]. The large $U$ in the Mo-layer splits the doubly valley-degenerate band into the lower and upper Hubbard bands[25] (Fig. 1c). The energy difference between the Mo-Hubbard bands and the dispersive W-band is further tunable by an electric field ($E$) perpendicular to the sample plane, which tunes the sublattice/interlayer potential difference[19-21]. The system realized a tunable Anderson lattice model[26-32].

Recent experiments on MoTe$_2$/WSe$_2$ bilayers have demonstrated the emergence of both nontrivial band topology[19, 21] and Kondo lattice[20, 33] physics. In particular, a topological phase transition from a charge-transfer insulator to a Chern insulator is observed[19] when the E-field inverts the W-band with the lower Hubbard band of the Mo-layer at a total hole filling factor $\nu = 1$; another transition from a band insulator to a topological insulator (TI) protected by spin-$S_z$ conservation has also been demonstrated[19] when the W-band inverts with the upper Hubbard band at $\nu = 2$. These results demonstrate a topologically nontrivial interlayer hopping $V$ in the material. In addition to the nontrivial band topology, gate tunable heavy fermions and magnetic field-induced Kondo breakdowns have also been demonstrated when the material is prepared in the Kondo lattice regime[20], in which the Mo-layer is kept in a triangular lattice Mott insulating state with a filling factor $\nu_f \approx 1$ and the W-layer is doped with itinerant holes at a filling factor $\nu_c < 1$. The itinerant holes are coupled to a triangular lattice of local magnetic moments in the Mott insulator via a topologically nontrivial Kondo exchange interaction (due to the topological nature of $V$), which has been predicted[26-28] to stabilize a 2D TKI at $\nu_f \approx \nu_c \approx 1$ and $\nu = \nu_f + \nu_c = 2$.

To search for the TKI, we fabricate dual-gated Hall bar devices of angle-aligned MoTe$_2$/WSe$_2$ moiré bilayers (Fig. 1d). The top and bottom gates allow independent control of $\nu$ and $E$. Compared to earlier transport studies[19-21] on the same material, we have used thinner (about 3 nm) hexagonal boron nitride (hBN) top gate dielectric, which allows the application of higher E-fields ($E > 1$ V/nm) without dielectric breakdown and enables us to access the $\nu = \nu_f + \nu_c = 1 + \nu_c$ Kondo lattice region from $\nu_c = 0$ to $\nu_c > 1$. In this study, we carry out transport measurements in three different geometries in the Hall bar device: the local, bulk, and nonlocal geometries (Fig. 1e-g). The local and nonlocal geometries[34, 35] are dominated by the helical edge state transport of the TKI whereas the bulk geometry[36] provides a proxy for the bulk resistance with substantially reduced edge state contributions. See Methods for details on device fabrication, electrical transport, and compressibility measurements.

**Electrostatics phase diagram**
We first examine the electrostatics phase diagram of the material. Figures 2a and 2b show the four-terminal longitudinal resistance $R_{xx}$ measured with the local geometry as a function of $\nu$ and $E$ under a perpendicular magnetic field $B_\perp = 0$ T and 14 T, respectively. The sample temperature is at $T = 1.6$ K unless otherwise specified. The grey-shaded regions are inaccessible due to dielectric breakdown at high E-fields and the poor electrical contacts at low E-fields. We also show in Fig. 2c the electrostatics phase diagram constructed from the data in Fig. 2a and 2b. Consistent with earlier studies[19-21], four distinct regions can be identified; dashed lines label their phase boundaries. The corresponding band alignment for each representative region is also shown in Fig. 2d.

The region below the lower orange boundary corresponds to a hole-doped Mo-layer only, i.e. only the Mo-Hubbard bands are doped with $\nu = \nu_f + 0$. Here a Mott insulator and a band insulator with divergent $R_{xx}$ are observed at $\nu = 1$ and $\nu = 2$, respectively; Wigner-Mott insulators at fractional fillings (e.g. $\nu = 1/2$ and $\nu = 2/3$) are also observed. On the other hand, the region above the upper orange boundary corresponds to a hole-doped W-layer only, i.e. only the dispersive W-band is doped with $\nu = 0 + \nu_c$. Here a metallic state is observed in the accessible region under $B_\perp = 0$ T; no correlated insulating state is observed due to the weak correlation effect here. At $B_\perp = 14$ T, a series of spin-valley-polarized Landau levels emerges, as revealed by the vertical stripes of Shubnikov-de Haas (SdH) oscillations; temperature dependence studies determine a hole effective mass about $0.5 m_0$ in the W-layer ($m_0$ is the free electron mass, see Extended Data Fig. 3).

The doped holes are shared between the two TMD layers in the remaining regions of the phase diagram. The physics at generic fillings $\nu_f$ and $\nu_c$ is beyond the scope of this study. However, the region bound by the two blue boundaries corresponds to the above-mentioned Kondo lattice region with $\nu = \nu_f + \nu_c = 1 + \nu_c$, i.e. the lower Hubbard band is fully filled and the W-band is partially filled (Fig. 2d). Whereas no clear phase boundaries can be identified at $B_\perp = 0$ T because of the Kondo coupling in this region, clear phase boundaries can be identified at $B_\perp = 14$ T, where Kondo singlets are broken down by the Zeeman field and SdH oscillations in the W-layer emerge[20] (see below for more discussions on magnetic Kondo breakdown). A doubling in the Landau level degeneracy is observed at $x \gtrsim 0.5$ as the Fermi energy exceeds the spin-/valley-Zeeman splitting in the W-layer[37]. The Mott gap size in the Mo-layer, as measured by the E-field span in the $\nu = 1 + \nu_c$ region[20], also shrinks with increasing $\nu_c$ due to free carrier screening of $U$ from the W-layer. Moreover, the $\nu = 1 + 1$ resistance peak disappears at $B_\perp = 14$ T, where the state becomes metallic. This is in contrast to the persistence of the $\nu = 2$ insulating state at E-fields below the Kondo lattice region (Fig. 2b), where the system is in the mixed valence regime with $\nu_f > 1$ and $\nu_c < 1$.

In the following, we will first demonstrate the emergence of Kondo lattice physics in the $\nu = 1 + \nu_c$ region. We will then present experimental evidence of a TKI at $\nu = 1 + 1$ through combined transport and compressibility measurements; we will also compare the results with the $\nu = 0 + 1$ state, in which no Kondo physics is expected. Finally, we will study the adiabatic evolution from a mixed valence TI (with $\nu_f > 1$ and $\nu_c < 1$) to a TKI (with $\nu_f \approx \nu_c \approx 1$) by continuously tuning $E$ at fixed $\nu = 2$.

**Kondo lattice physics at $\nu = 1 + \nu_c$**

Figure 3a shows the temperature dependence of the local resistance $R_{xx}$ at varying $\nu_c$ along the arrow in Fig. 2a ($B_\perp = 0$). Fermi liquid behavior $R_{xx} = A \times T^2 + R_0$ is observed at $\nu_c < 1$ for temperatures below a coherence temperature scale $T^*$ ($R_0$ is the disorder-limited residual resistance). As shown in Fig. 3b, $T^*$ first increases with $\nu_c$ and saturates to about 30 K in the $\nu_c \to 1$ limit (see Methods for the determination of $T^*$). Meanwhile, the Kadowaki-Woods coefficient[38] $A$ decreases with $\nu_c$ in a way approximately following $T^* \propto A^{-0.5}$. In contrast to the Fermi liquid behavior at $\nu_c < 1$, $R_{xx}$ at $\nu_c = 1$ first follows a metallic temperature dependence for $T \gtrsim 20$ K, and for $T \lesssim 20$ K $R_{xx}$ increases and saturates in the low-temperature limit to a value (16 kΩ) comparable to $\frac{h}{2e^2}$ ($h$ and $e$ denote the Planck constant and electron charge, respectively).

We also examine $R_{xx}$ and $R_{xy}$ (Hall resistance) as a function of $\nu_c$ and $B_\perp$ in Fig. 3c and 3d. Line cuts of $R_{xx}$ and $R_H$ (Hall coefficient) at selected $\nu_c$'s are shown in Fig. 3e and 3f. With increasing $B_\perp$, $R_{xx}$ first increases and reaches a maximum near a characteristic field $B_{\perp C}$, beyond which $R_{xx}$ drops substantially and SdH oscillations emerge. Simultaneously, $R_H$ (or $R_{xy}$) changes sign at $B_{\perp C}$ and "jumps" by a step approximately given by the moiré density $n_M$. Similar to $T^*$, $B_{\perp C}$ increases with $\nu_c$ and saturates to about 11 T in the $\nu_c \to 1$ limit (Fig. 3b-d).

The results above demonstrate the emergence of Kondo lattice physics at $\nu = \nu_f + \nu_c = 1 + \nu_c$. In particular, Kondo singlets emerge at $T \lesssim T^*$; the coherent scattering of the itinerant holes in the W-layer with the lattice of local magnetic moments in the Mo-layer gives rise to a heavy Fermi liquid[20] with enhanced quasiparticle effective mass $m^* \propto A^{0.5}$, which is expected to scale with $T^*$ as $m^* \propto A^{0.5} \propto \frac{1}{T^*}$ (Ref. [39]). The increase in $T^*$ with $x$ is consistent with the predicted strengthening of the effective Kondo exchange interaction with increasing itinerant hole density[29, 32, 39]. In addition to the thermal breakdown of the Kondo singlets at $T \gtrsim T^*$, Kondo breakdown[40] can also occur under an external Zeeman field $B_\perp$ when the associated Zeeman energy increases beyond $k_B T^*$ and forces the itinerant hole spin to align with the local moment spin ($k_B$ is the Boltzmann constant). This results in the breakdown of the heavy fermions and therefore the immediate appearance of SdH oscillations from the itinerant holes at $B_\perp > B_{\perp C}$. The breakdown of the heavy fermions also gives rise to a size jump[40] in the Fermi surface from $1 + \nu_c$ to $\nu_c$ and therefore the observed jump in $R_H$. These results are fully consistent with earlier reports of emergent heavy fermion physics in MoTe$_2$/WSe$_2$ moiré bilayers[20].

**Evidence of TKI at $\nu = 1 + 1$**

We now examine the $\nu = 1 + 1$ state. Figure 4a shows the local $R_{xx}$ as a function of $\nu_c$ at $T = 1.6$ K; the temperature dependence at $\nu_c = 1$ is summarized in Fig. 4b. At $B_\perp = 0$ T, a $R_{xx}$ peak is observed at $\nu = 1 + 1$; with decreasing temperature below $T \approx 20$ K, the peak value increases and plateaus to about 16 kΩ (comparable to $\frac{h}{2e^2}$) in the low-temperature limit. In contrast, the $R_{xx}$ peak disappears after the Kondo breakdown at $B_\perp = 14$ T; it also disappears in the $\nu = 0 + \nu_c$ regime, where the Mott-localized holes in the Mo-layer are depleted. In both cases, $R_{xx}$ at $\nu_c = 1$ show metallic temperature dependence over the examined temperature range.

We access the bulk transport of the material by the bulk geometry in Fig. 1g. In contrast to the local $R_{xx}$, the bulk resistance $R_{bulk}$ shows a much higher peak value at $\nu = 1 + 1$ (Fig. 4a); the peak value also increases quickly with decreasing temperature, demonstrating an insulating behavior (Fig. 4b). We further access the bulk incompressibility of the sample by penetration capacitance measurements (Fig. 4c). An incompressibility peak at $\nu = 1 + 1$ is observed at $B_\perp = 0$ T; integration of the peak area gives a bulk charge gap about 1 meV for the incompressible state. In contrast, both the $\nu = 1 + 1$ state at $B_\perp = 14$ T and the $\nu = 0 + 1$ state are compressible, consistent with the metallic transport in Fig. 4b.

Next, we examine the nonlocal transport properties. We bias current $I_{46}$ between the source and drain electrodes (4-6, see Fig. 1f) and measure the nonlocal voltage drop across the electrode pairs 1-9 ($V_{19}$), 1-2 ($V_{12}$) and 9-8 ($V_{98}$). The nonlocal resistances $R_{19,46} \equiv \frac{V_{19}}{I_{46}}$, $R_{12,46}$ and $R_{98,46}$ as a function of $\nu_c$ is shown in Fig. 4d. A negligible nonlocal resistance is observed for the metallic states away from $\nu = 1 + 1$ due to the exponential suppression of nonlocal bulk transport with distance away from the source-drain pair 4-6. On the other hand, a large nonlocal resistance peak is observed at $\nu = 1 + 1$ in all cases. In particular, the $R_{19,46}$ peak is about twice the $R_{98,46}$ peak; the $R_{98,46}$ peak is of similar height to $R_{12,46}$ but of opposite sign.

We further study the anisotropic magnetoresistance near $\nu = 1 + 1$. Figure 4e shows the local $R_{xx}$ as a function of $\nu_c$ at varying out-of-plane ($B_\perp$) and in-plane ($B_\parallel$) magnetic fields. Whereas the $\nu = 1 + 1$ state shows negligible magnetoresistance under $B_\perp$, it shows a large positive magnetoresistance under $B_\parallel$ and exhibits a cusp-shape dependence on $B_\parallel$ (Fig. 4f; see temperature dependence in Extended Data Fig. 4). Moreover, the large in-plane magnetoresistance disappears at the metallic states away from $\nu = 1 + 1$ (e.g. $\nu = 1 + 0.8$). Note that the magnetic field here is substantially lower than the Kondo breakdown field $B_{\perp C} \approx 11$ T at $\nu = 1 + 1$ (Fig. 3). For $B_\perp > B_{\perp C}$, the $\nu = 1 + 1$ state is metallic (as shown above) with a Hall density $n_H \approx n_M$.

The results in Fig. 4 demonstrate the emergence of a TI with helical edge states protected by spin-$S_z$ conservation[41, 42] in the Kondo lattice region (i.e. the emergence of a TKI). First, both the bulk resistance and incompressibility measurements show an insulating state at $\nu = 1 + 1$. Yet the local $R_{xx}$ shows a nearly quantized value at $\frac{h}{2e^2}$ at low temperatures, suggesting helical edge state transport. The ≈20% higher $R_{xx}$ than $\frac{h}{2e^2}$ suggests that the coherence length of the helical edge states is comparable to or shorter than the separation between adjacent electrodes (about 2 μm). Such imperfect quantization is common for helical edge states that are susceptible to backscattering from magnetic impurities, thermal excitations, and electron-electron interactions[43-48].

The presence of helical edge state transport is further supported by the nonlocal transport results. Because the helical edge states can equilibrate with each contact in the device and emanate at the contact chemical potential, an equivalent circuit (inset of Fig. 4d) for the nonlocal geometry can be derived using the Landauer-Buttiker formalism[34] (Methods). This circuit model naturally explains the observed $R_{19,46} \approx 2R_{98,46}$ and $R_{98,46} \approx -R_{12,46}$ at $\nu = 1 + 1$; in particular, $R_{98,46} \approx -R_{12,46}$ uniquely demonstrates edge state transport as illustrated by the edge currents in the inset of Fig. 4d. ($R_{98,46}$ and $R_{12,46}$ would be of the same sign if they were dominated by bulk transport.) The model also predicts $R_{19,46} = \frac{2h}{5e^2} \approx 10.3$ k$\Omega$ and $R_{98,46} \approx -R_{12,46} = \frac{h}{5e^2} \approx 5.2$ k$\Omega$; these values are in reasonable agreement with the measured values at $\nu = 1 + 1$. The discrepancies likely come from the remnant bulk transport in the material and/or backscattering in the edge states; the former (latter) lowers (raises) the nonlocal resistance from the expected values.

Next, the anisotropic magnetoresistance data in Fig. 4e,f are fully consistent with helical edge state transport protected by spin-$S_z$ conservation[41, 42]. In particular, spin-$S_z$ remains a good quantum number under $B_\perp$; helical edge state transport remains robust (until $B_\perp$ approaches $B_{\perp C}$); a negligible magnetoresistance is thus observed at $\nu = 1 + 1$ for small $B_\perp$. In contrast, $B_\parallel$ destroys spin-$S_z$ conservation and induces backscattering for the edge state transport, resulting in a large positive magnetoresistance[36, 41, 42] even for small $B_\parallel$. Yet the bulk magnetoresistance for the compressible states (e.g. $\nu = 1 + 0.8$) is negligible under $B_\parallel$ because of the strong Ising spin-orbit coupling in the material[18]. The results clearly demonstrate the distinct transport properties for the bulk and edge states.

Lastly, the high-temperature metallic transport and the low-temperature quantum spin Hall transport observed for $R_{xx}$ (Fig. 3a and 4b) support the picture of a TKI induced by the Kondo screening of local moments below the Kondo temperature $T^*$. The relevance of Kondo interactions is also supported by the disappearance of the insulating state at $\nu = 0 + 1$ when the local moments are depleted in the Mo-layer. Moreover, the transition from the TKI (with nearly quantized $R_{xx}$) to a metallic state (with small $R_{xx}$ and $n_H \approx n_M$) near $B_\perp = B_{\perp C}$ (Fig. 3) demonstrates a magnetic breakdown of the Kondo singlets. The breakdown happens when the Zeeman energy scale $g\mu_B B_{\perp C}/k_B \approx 60$ K becomes comparable to $T^* \approx 30$ K ($g \approx 10$ and $\mu_B$ is the Bohr magneton); it is continuously connected to the nearby Kondo breakdowns for the $\nu_c < 1$ heavy fermion phase (Fig. 3c,d). In other words, the TKI at $\nu = 1 + 1$ can be considered as a parent state for the heavy fermion phase at generic $\nu_c < 1$ (Ref. [3]).

**Mixed-valence to Kondo lattice crossover**
Finally, we examine the adiabatic crossover from the TKI ($\nu_f \approx \nu_c \approx 1$) to a mixed-valence TI[26] ($\nu_f > 1$ and $\nu_c < 1$) by tuning $E$ at fixed $\nu = 2$ (Fig. 1c). Figure 5a shows the E-field dependence of the local $R_{xx}$ at varying temperatures. The $\nu = 2$ state is in the Kondo lattice regime for $E > E_2 \approx 0.73$ V/nm, in the mixed valence regime for $E_1 \approx 0.57 > E > E_2 \approx 0.73$ V/nm, and is a trivial band insulator for $E \lesssim E_1$. The phase boundaries are determined by the $R_{xx}$ map in Fig. 2b. Whereas $R_{xx}$ in the Kondo lattice regime shows weak temperature dependence in the low-temperature limit and saturates to a value up to about 2-3 times of $\frac{h}{2e^2}$, $R_{xx}$ near the middle of the mixed valence regime increases substantially at low temperatures and reaches a value many times of $\frac{h}{e^2}$. Only at lower E-fields, where the $\nu = 2$ state is close to a transition to the band insulator, the behavior becomes similar to that in the Kondo lattice regime.

We examine the electronic incompressibility of the $\nu = 2$ state in the mixed valence regime (Extended Data Fig. 5). The $\nu = 2$ state is in general incompressible except near $E = E_1$. Extended Data Fig. 5b shows the E-field dependence of the extracted charge gap of the $\nu = 2$ state. The charge gap closes at the boundary ($E \approx E_1$) separating the band insulator and the mixed valence regime. Earlier studies[19] have shown that this gap closure corresponds to a topological phase transition from a band insulator to a TI induced by an E-field-tuned inversion between the W-band and the Mo-upper Hubbard band (Fig. 1c). In contrast to the gap closure near $E = E_1$, the charge gap evolves smoothly from the mixed valence regime to the Kondo lattice regime near $E = E_2$.

To demonstrate that the $\nu = 2$ state is a TI throughout the mixed valence regime ($E_1 \gtrsim E \gtrsim E_2$), we study the nonlocal resistance $R_{19,46}$ as a function of $\nu$ and $E$ in Fig. 5b. A large nonlocal resistance is observed at $\nu = 2$ for $E \gtrsim E_1$, consistent with the emergence of helical edge states. The presence of helical edge states protected by spin-$S_z$ conservation is further demonstrated by the in-plane magnetoresistance $\frac{R_{xx}(B_\parallel=0\text{T})}{R_{xx}(B_\parallel=1\text{T})}$ as a function of $\nu$ and $E$ in Fig. 5c. A large positive in-plane magnetoresistance is observed only near $\nu = 2$ for $E \gtrsim E_1$. Similar to the TKI, the in-plane magnetoresistance exhibits a cusp-shape and the out-of-plane magnetoresistance is negligible in the mixed valence regime (Fig. 5d).

The results above demonstrate the emergence of a mixed-valence TI at $\nu = 2$ for $E_1 \gtrsim E \gtrsim E_2$. Unlike the gap closure at the topological phase transition between the band insulator and the mixed valence TI near $E = E_1$, the smooth evolution of the charge gap from the mixed valence to the Kondo lattice regime shows that the mixed valence TI and the TKI are adiabatically connected; only the ratio $\frac{v_f}{v_c}$ varies across the crossover, consistent with theoretical expectations[26]. Compared to the TKI, the helical edge states of the mixed valence TI have much shorter coherence lengths due to enhanced back scattering[46, 47, 49], as illustrated by the much higher $R_{xx}$ and $R_{19,46}$ than $\frac{h}{e^2}$ in Fig. 5a and 5b, respectively. Moreover, $R_{xx}$ of the mixed valence TI is highly sensitive to the transport bias current; scaling analyses show evidence of one-dimensional (1D) Luttinger liquid physics[48, 50-52] in the mixed valence regime (Extended Data Fig. 6). The origin of the much shorter coherence length and the associated Luttinger liquid physics for the mixed-valence TI deserves further investigation.

**Conclusion**
In conclusion, by combining electrical transport and compressibility measurements, we demonstrate experimental evidence of a TKI with helical edge state transport protected by spin-$S_z$ conservation at $\nu = 1 + 1$ in the Kondo lattice region. Temperature and magnetic field dependence studies support the emergence of Kondo singlets for $T \lesssim T^*$ and $B_\perp <$

$B_{\perp C}$. Our results pave the path for further exploration of tunable topological Kondo physics in moiré materials.

## Methods
### Device fabrication
We fabricated dual-gated Hall bar devices of angle-aligned MoTe$_2$/WSe$_2$ using the layer-by-layer dry transfer method detailed in previous studies[53, 54]. The constituent atomically thin flakes were exfoliated from bulk crystals onto Si substrates with a 285 nm SiO$_2$ layer and identified by optical reflection contrast. The flakes were picked up sequentially using a polycarbonate (PC) stamp to form the heterostructure shown in Fig. 1d. The crystallographic orientations of the MoTe$_2$ and WSe$_2$ monolayers and their twist angle (0-degree aligned or 60-degree aligned) were determined by angle-resolved optical second harmonic generation (SHG)[55, 56] with about $\pm 0.5°$ uncertainty. All the processes involving MoTe$_2$ before encapsulation by hexagonal boron nitride (hBN) were conducted in a nitrogen-filled glovebox to minimize sample oxidation.

To produce high-quality devices for transport studies, we divided the stacking process into several steps. We first released a bottom gate of few-layer graphite and hBN ($\approx$ 10 nm) onto a Si/SiO$_2$ substrate. Thin Pt electrodes ($\sim$ 8 nm) were patterned into a Hall bar geometry on the hBN by electron-beam lithography and evaporation. Polymer residues on the bottom gate were cleaned by atomic force microscopy (AFM) in contact mode with a typical force of 500 nN. The remaining stack, consisting of the MoTe$_2$/WSe$_2$ hetero-bilayer and a hBN/graphite top gate, was released onto the pre-patterned Pt electrodes at 200°C. A geometry with a narrower top graphite gate and a wider bottom graphite gate allows us to study the electrical transport properties of the top gate-defined region only without contributions from other parallel channels. Compared to earlier studies[19, 20], a thinner hBN top gate dielectric (< 4 nm) was used to achieve a larger breakdown electric field ($\approx$ 1.4 V/nm). We have studied a total of five devices, all of which show similar results (see Extended Data Fig. 1 for data from another device).

### Electrical measurements
The electrical transport measurements were performed in a closed-cycle $^4$He cryostat equipped with a 14 T superconducting magnet (Oxford TeslatronPT). Standard low-frequency (13.77 Hz) lock-in techniques were employed to measure the four-terminal resistance with a 1 mV bias excitation at the source electrode. Both the voltage drop at the probe electrodes and the source-drain current (below 100 nA) were recorded. Different measurement geometries, as shown in Fig. 1e-g, were used to measure the longitudinal and Hall resistances, the nonlocal resistance, and the bulk resistance. Voltage amplifiers with large input impedance (100 MΩ) were used to measure the sample resistance up to about 10 MΩ. All data were taken at 1.6 K unless otherwise specified.

### Compressibility measurements
The compressibility measurements were performed in the same Hall bar device and cryostat as the electrical transport measurements. Details have been reported in previous studies[19, 57]. A 5 mV excitation was applied to the bottom gate and the displacement current was collected from the top gate through a high electron mobility transistor (HEMT) to measure the penetration capacitance $C_p$ with the MoTe$_2$/WSe$_2$ bilayer grounded. A commercial HEMT (FHX35X) was mounted vertically on the same chip near the sample as the first-stage amplifier[58, 59] to eliminate parasitic capacitance. Standard lock-in techniques with a modulation frequency of 437.77 Hz were employed to measure the differential capacitance.

Based on a lumped circuit model, the penetration capacitance can be written as

$$\frac{C_p}{C_{series}} = \frac{1}{1+\frac{C_Q}{C_{parallel}}}.$$

Here $C_Q = e^2 dn/d\mu$ is the quantum capacitance, $C_{series} \equiv C_t C_b/(C_t+C_b)$ and $C_{parallel} \equiv C_t+C_b$ are, respectively, the series and parallel combinations of the geometrical top-gate ($C_t$) and bottom-gate ($C_b$) capacitances. The thermodynamic gap size of an insulating state can be obtained as $\Delta\mu = \int (e^2/C_b)dn = e\int (C_p/C_b)dV_{tg}$.

**Determination of $T^*$ and $A$**

The Kondo temperature $T^*$ describes a crossover temperature scale that separates the coherent and incoherent transport at $T < T^*$ and $T > T^*$, respectively[60]. Above $T^*$, the unscreened local moments provide perturbative Kondo scattering to the itinerant electrons, resulting in a weakly temperature-dependent resistance. At temperatures below $T^*$, the lattice of local moments is coherently screened, resulting in a quick drop in the sample resistivity with decreasing temperature and the emergence of a coherent Landau Fermi liquid.

In our experiment, $T^*$ was estimated as a resistance bump or a change in the resistance slope in Fig. 3a, serving as a measure of the effective Fermi temperature of the correlated electrons. We extracted the crossover temperature $T^*$ by identifying the intersection of two fitted lines representing the resistance in the coherent and incoherent transport regimes (Extended Data Fig. 2). As coherent quasiparticles emerge below $T^*$, Landau Fermi liquid behavior $R_{xx}(T) = R_0 + AT^2$ was observed at low temperatures. The coefficient $A$ was extracted by fitting the low-temperature part of the temperature-dependent resistance (Extended Data Fig. 2) to provide a measure for the quasiparticle effective mass[38].

**Analysis of nonlocal transport measurements**

Nonlocal transport measurements were performed to demonstrate the presence of helical edge states. As shown in Fig. 1f and the inset of Fig. 4d, the source-drain pair and the voltage probe pair are widely separated to minimize the bulk transport contribution[34, 35]. In our device, the main device channel has a dimension of $14 \times 2 \ \mu m^2$; the separation between adjacent electrodes is $2 \ \mu m$. As observed in Fig. 4d, the diffusive bulk transport away from $\nu = 2$ results in a negligible contribution to the nonlocal resistances; only the extended edge channels at $\nu = 2$ contribute to the nonlocal signals.

The nonlocal edge state transport can be well described by the Landauer-Büttiker formulism[61]. In contrast to the chiral edge states in the quantum Hall regime, the nonlocal resistance is in general finite for a 2D TI because of its counter-propagating edge channels. These channels propagate at different chemical potentials inherited from the respective outlet contacts, forming a net current flow. Despite the non-dissipative nature of the helical edge states, quantum phase coherence is lost at the contacts because the counter-propagating edge channels are forced to equilibrate at the same potential, resulting in a resistance of $h/e^2$. Consequently, the nonlocal measurement circuit can be simplified to an equivalent resistance network, as shown in the inset in Fig. 4d, in which each resistor carries a resistance quantum.

In Fig. 4d, three different configurations were used to demonstrate the reliability of the nonlocal measurements and the existence of edge states. Using the equivalent circuit model, we can obtain $R_{19,46} = 2h/5e^2 \approx 10.3 \text{ k}\Omega$ and $R_{98,46} \approx -R_{12,46} = h/5e^2 \approx 5.2 \text{ k}\Omega$. These values are in reasonable agreement with the measured values at $\nu = 1 + 1$. Note that additional phase decoherence for the edge states may also occur in charge puddles induced by sample inhomogeneity near the edge, effectively acting as dephasing centers and leading to deviations from the expected quantized values[44-46].

**Acknowledgments**
We thank Debanjan Chowdhury, Daniele Guerci, Andrew Millis, Jedediah Pixley and Qimiao Si for their insightful discussions.

# Figures

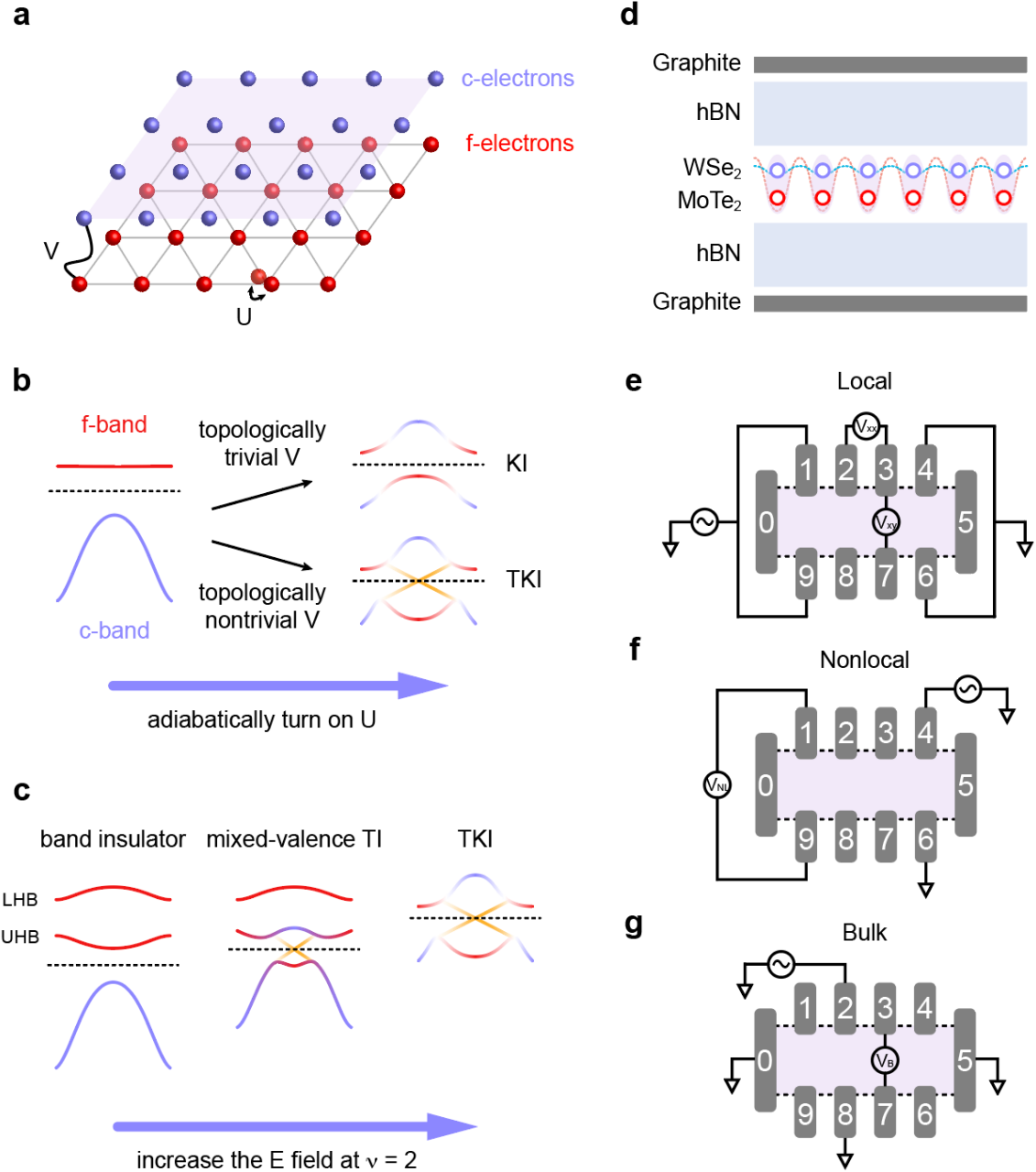

**Figure 1 | Moiré Kondo lattice in MoTe$_2$/WSe$_2$ devices. a,** Schematic bilayer Anderson lattice model, in which the c-electrons support a dispersive energy band, the f-electrons support a flat band, and the two bands are coupled by a hybridization term $V$. $U$ is the on-site Coulomb repulsion for double occupancy in the f-band. **b,** Schematic evolution of the band structure from a band insulator ($\nu_f = 2$ and $\nu_c = 0$) to a TKI (KI) with a topologically nontrivial (trivial) hybridization term $V$ as the on-site Coulomb repulsion is adiabatically turned on. The dashed line denotes the Fermi level. **c,** Electric field-tuned band structure at $\nu = 2$. A topological phase transition from a band insulator ($\nu_f = 2$ and $\nu_c = 0$) to a mixed-valence TI ($\nu_f > 1$ and $\nu_c < 1$) occurs when the WSe$_2$ c-band inverts with the MoTe$_2$ upper Hubbard band (UHB). Then an adiabatic crossover from a mixed-valence TI

to a TKI ($v_f \approx v_c \approx 1$) follows at high fields. **d,** Schematic cross-section of a dual-gated Hall bar device of MoTe$_2$/WSe$_2$ moiré bilayers. The Kondo coupling between the lattice of local moments in the Mo-layer and the itinerant carriers in the W-layer forms a lattice of Kondo singlets and stabilizes a TKI. **e-g,** Electrical connections for the local (**e**), nonlocal (**f**), and bulk (**g**) measurement geometries. $V_{xx}$ and $V_{xy}$ in **e** denote the longitudinal and transverse voltage drops, respectively; $V_{NL}$ in **f** denotes the nonlocal voltage drop; $V_B$ in **g** denotes the voltage drop in the bulk.

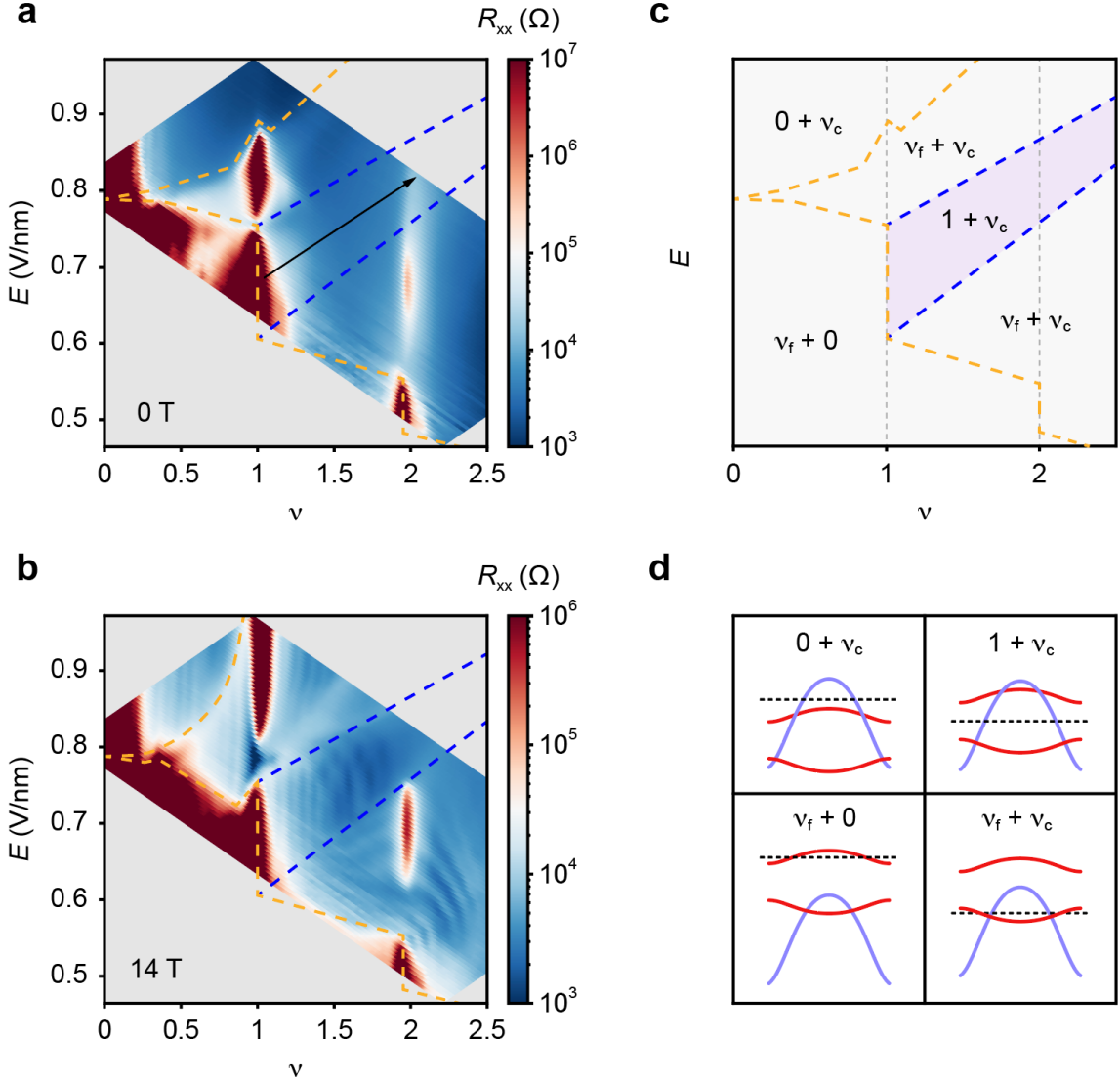

**Figure 2 | Electrostatics phase diagram. a,b,** The local longitudinal resistance $R_{xx}$ at $T = 1.6$ K as a function of $\nu$ and $E$ at $B_\perp = 0$ T (**a**) and 14 T (**b**). The orange dashed lines separate the layer-polarized and the layer-hybridized regions. The Kondo lattice region with $\nu = 1 + \nu_c$ is enclosed by the blue dashed lines. The arrow in **a** denotes the filling scan direction within the Kondo lattice region in our experiment. **c,** Schematic electrostatics phase diagram of MoTe$_2$/WSe$_2$ moiré versus $\nu$ and $E$. The same dashed lines in **a** separate the different regions, which are labeled by the filling factors in the Mo- and W-layer ($\nu_f$ and $\nu_c$, respectively). **d,** Corresponding band alignment for each representative region in **c**. The hole doping density and the alignment between the MoTe$_2$ Hubbard bands (red) and the WSe$_2$ c-band are controlled by electrical gating. The dashed line denotes the Fermi level. The hybridization of the two bands is not shown for simplicity. (Results are from device 1.)

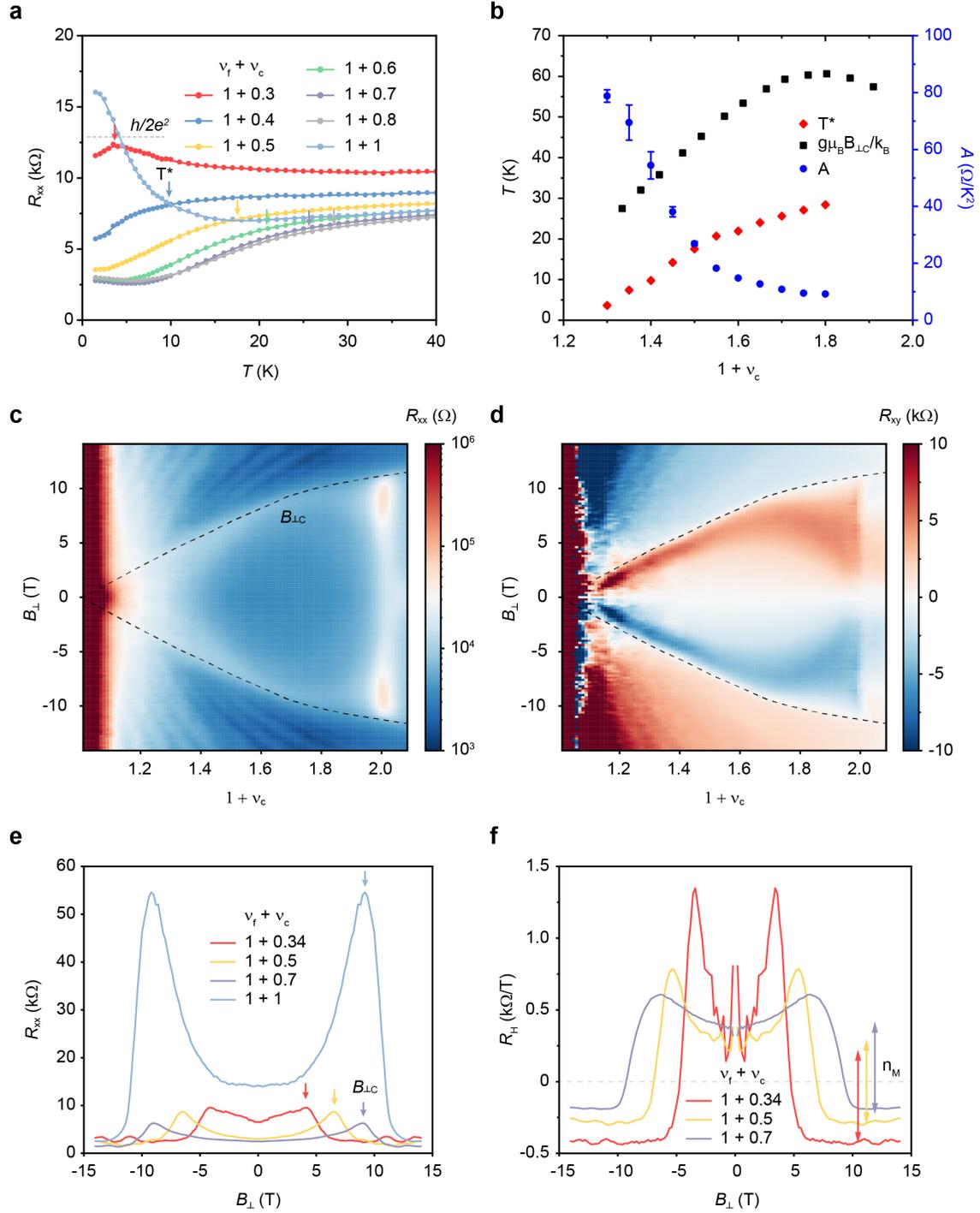

**Figure 3 | Kondo lattice physics at $\nu = 1 + \nu_c$. a,** Temperature dependence of $R_{xx}$ at varying $\nu_c$ along the arrow direction in Fig. 2a ($B_\perp = 0$ T). The arrows mark the Kondo temperature $T^*$ (see Methods for its determination). The grey dashed line marks the expected quantized value for a TI. **b,** The extracted $T^*$ (red), the critical Zeeman energy $g\mu_B B_{\perp C}/k_B$ (black) and the coefficient $A$ (blue) as a function of $\nu_c$ with $\nu_f = 1$. **c,d,** $R_{xx}$ (**c**) and $R_{xy}$ (**d**) as a function of $\nu = 1 + \nu_c$ and $B_\perp$ in the Kondo lattice region ($T = 1.6$ K). The dashed lines mark the Kondo breakdown critical field $B_{\perp C}$. SdH oscillations in $R_{xx}$

emerge at $B_\perp > B_{\perp C}$, where a sign change in $R_{xy}$ is simultaneously observed. **e,f,** Dependence of $R_{xx}$ (**e**) and the Hall coefficient $R_H$ (**f**) on $B_\perp$ at varying $\nu_c$ with $\nu_f = 1$ ($T = 1.6$ K). The arrows in **e** denote the critical field $B_{\perp C}$. The arrows in **f** denote the size of the jump in $R_H$ that corresponds to a jump in the Hall density by $n_M$ near $B_{\perp C}$. (Results are from device 1.)

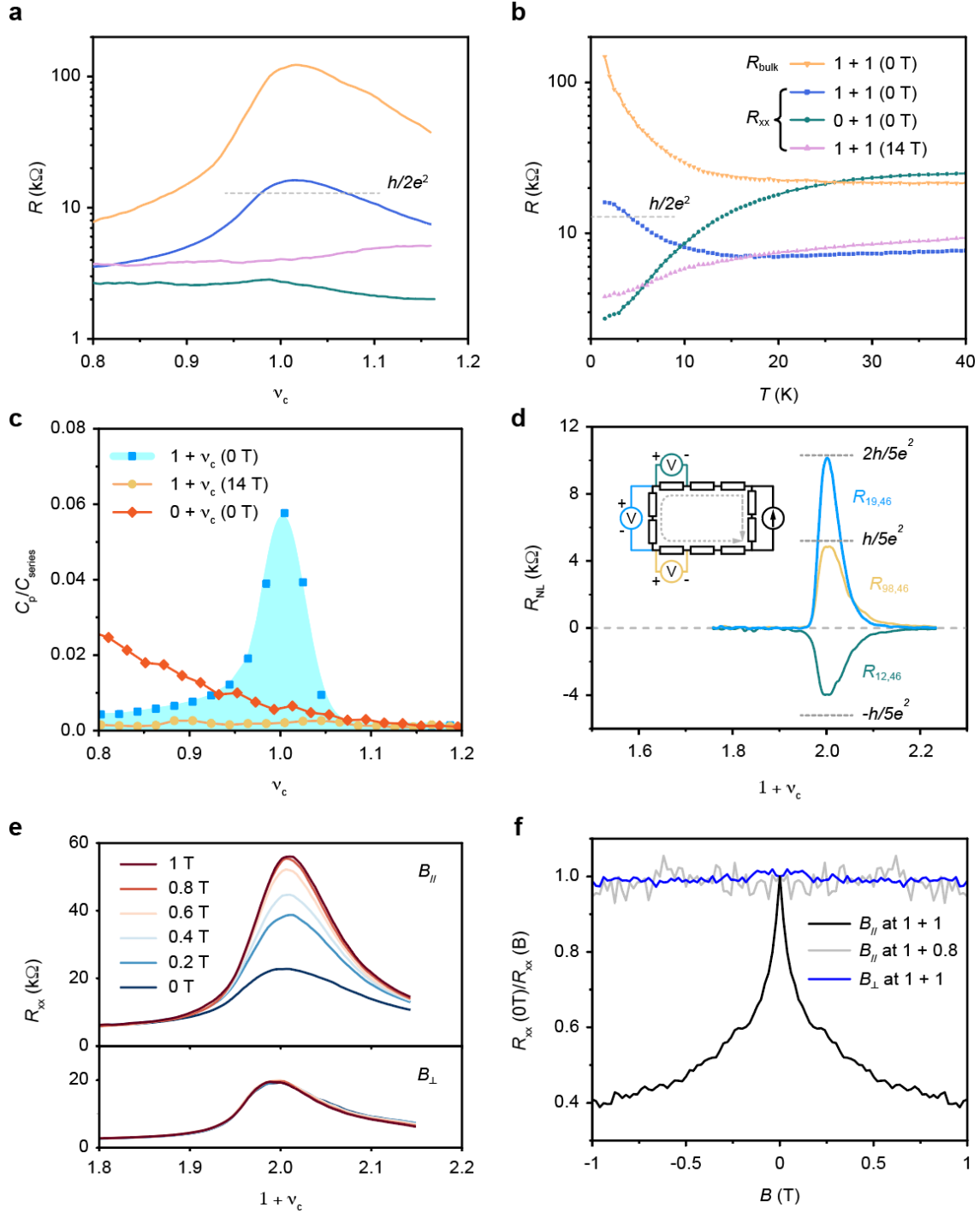

**Figure 4 | Topological Kondo insulator at $\nu = 1 + 1$. a,** Dependence of the sample resistance on $\nu_c$ at $T = 1.6$ K. $R_{xx}$ for $\nu = 1 + \nu_c$ at $B_\perp = 0$ T, $\nu = 1 + \nu_c$ at $B_\perp = 14$ T and $\nu = 0 + \nu_c$ at $B_\perp = 0$ T are shown in blue, pink, and green, respectively. The bulk resistance $R_{bulk}$ for $\nu = 1 + \nu_c$ at $B_\perp = 0$ T is shown in orange. **b,** Temperature dependence of the corresponding resistances at $\nu_c = 1$, shown by the same colors as in **a**. The dashed line marks $\frac{h}{2e^2}$. **c,** The normalized penetration capacitance $C_p/C_{series}$ at $T = 1.6$ K as a function of $\nu_c$ for $\nu = 1 + \nu_c$ at $B_\perp = 0$ T (blue) and 14 T (orange), and for $\nu = $

$0 + \nu_c$ at $B_\perp = 0$ T (red). The charge gap for the $\nu = 1 + 1$ state is proportional to the blue-shaded area. **d,** Dependence of the nonlocal resistance $R_{NL}$ on $\nu = 1 + \nu_c$ under different measurement configurations shown by the equivalent circuit diagram in the inset ($T = 1.6$ K). The different $R_{NL}$'s are labeled by the source-drain and voltage probe pairs (see main text). The horizontal dashed lines mark the expected quantized values at $\nu = 1 + 1$. The dashed arrows in the inset show the bias current flow along the sample edges at $\nu = 1 + 1$. **e,** Dependence of $R_{xx}$ on $\nu = 1 + \nu_c$ under varying $B_\parallel$ (top panel) and $B_\perp$ (bottom panel) at $T = 1.6$ K. **f,** In-plane (black and grey) and out-of-plane (blue) magnetoresistance $R_{xx}(0\,T)/R_{xx}(B)$ at $\nu = 1 + 1$ (black and blue) and $\nu = 1 + 0.8$ (grey). Strong magnetoresistance is observed only at $\nu = 1 + 1$ under in-plane fields. (The results in **a-c,e,f,** and **d** are from devices 1 and 2, respectively.)

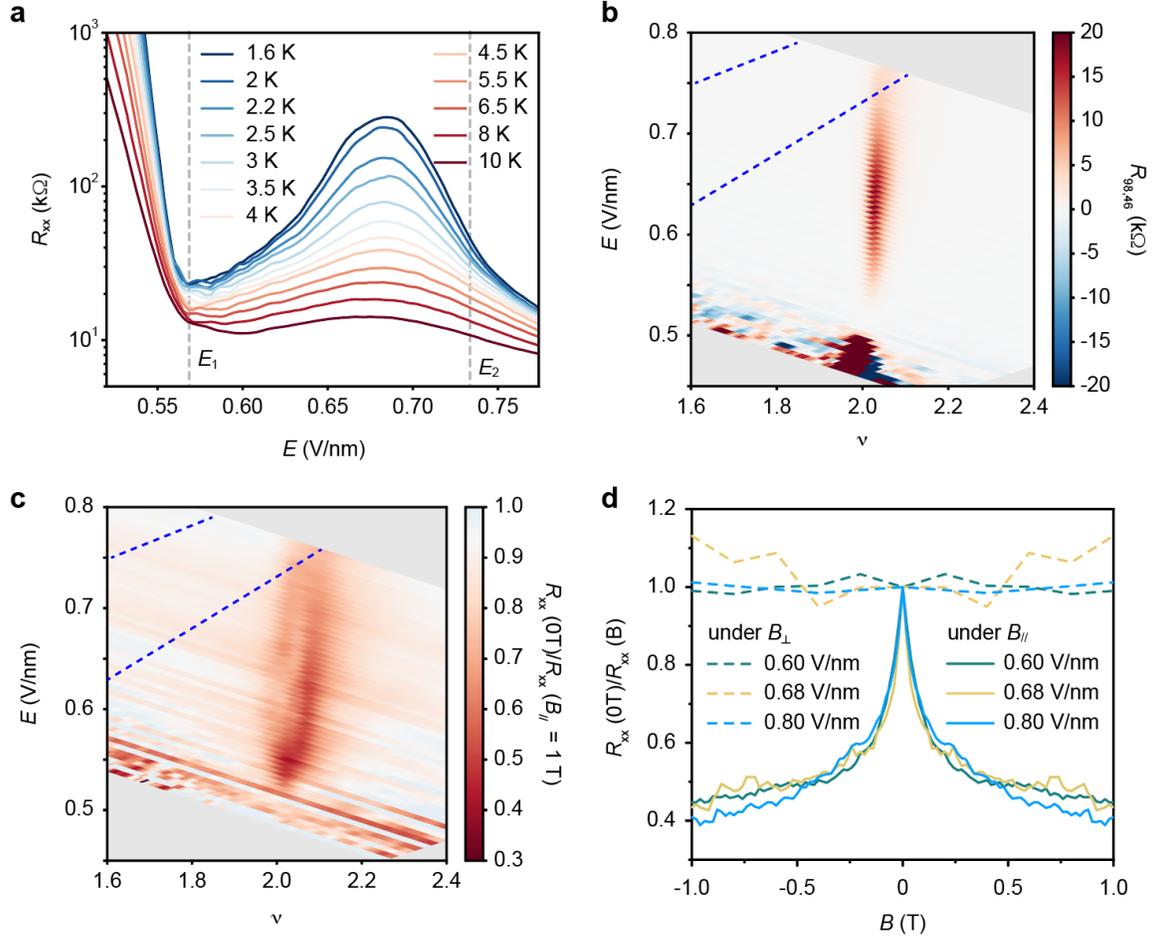

**Figure 5 | Electrical field dependence at $\nu = 2$. a,** Electric field dependence of $R_{xx}$ at varying temperatures ($B_\perp = 0$ T) at $\nu = 2$. The grey dashed lines indicate the phase boundaries ($E_1$ and $E_2$, see main text) separating three regimes: band insulator, mixed-valence TI and TKI. **b,c,** Nonlocal resistance $R_{98,46}$ (**b**) and in-plane magnetoresistance $R_{xx}$ (0 T)/$R_{xx}$ ($B_\parallel = 1$ T) (**c**) as a function of $\nu$ and $E$ at $T = 1.6$ K and $B_\perp = 0$ T. The Kondo lattice region is enclosed by blue dashed lines. An enhanced nonlocal signal and a suppressed in-plane magnetoresistance are observed throughout the mixed-valence TI and TKI regions (i.e. after band inversion). **d,** In-plane (solid) and out-of-plane (dashed) magnetoresistance $R_{xx}$ (0 T)/$R_{xx}$ (B) under varying $E$-fields at $\nu = 2$ ($T = 1.6$ K). (The results in **a,d,** and **b,c** are from devices 1 and 2, respectively.)

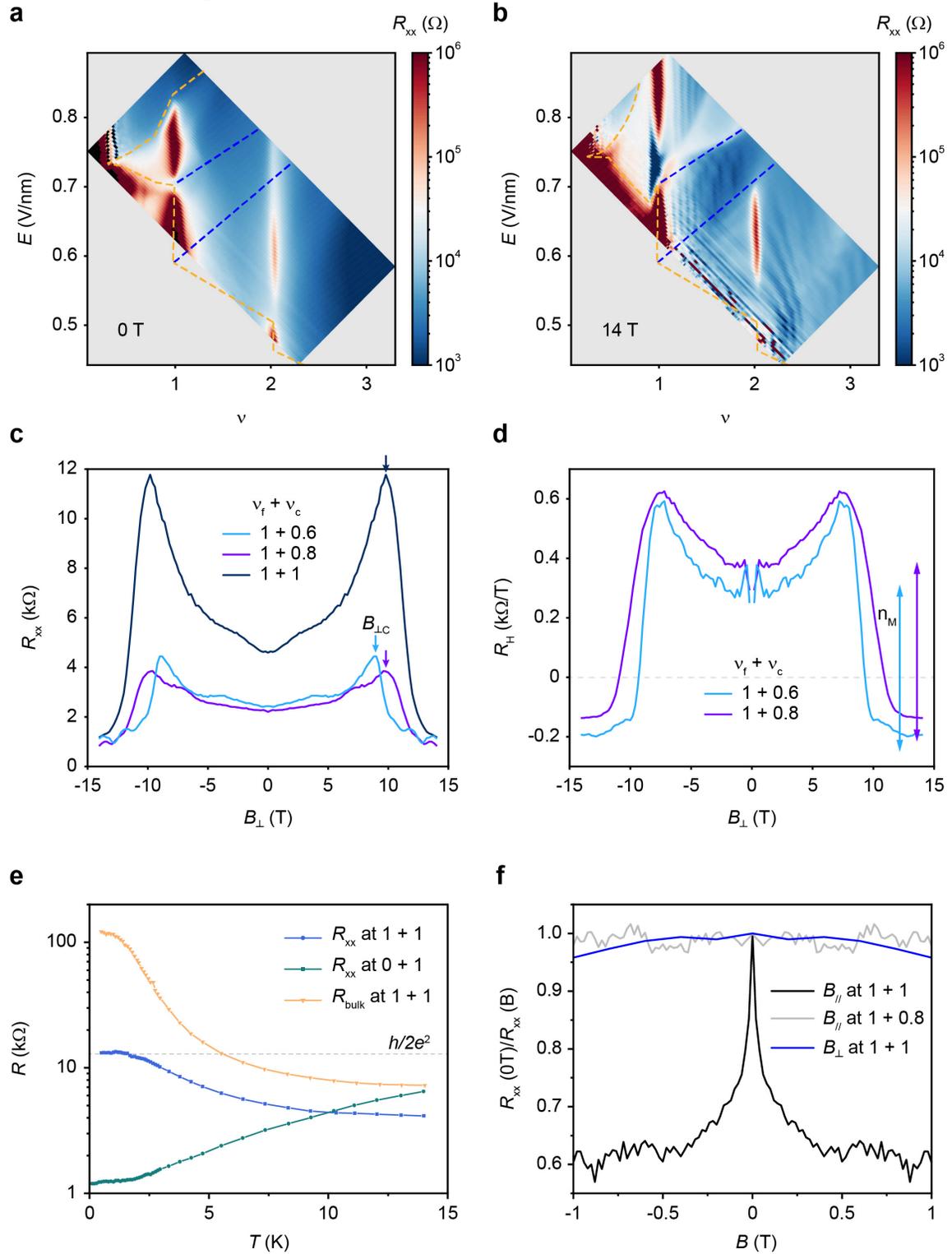

**Extended Data Figure 1 | Main results from device 2. a,b,** The local longitudinal resistance $R_{xx}$ at $T = 1.6$ K as a function of $\nu$ and $E$ at $B_\perp = 0$ T (**a**) and 14 T (**b**). The orange dashed lines separate the layer-polarized and the layer-hybridized regions. The Kondo lattice region with $\nu = 1 + \nu_c$ is enclosed by the blue dashed lines. **c,d,**

Dependence of $R_{xx}$ (**c**) and $R_H$ (**d**) on $B_\perp$ at varying $\nu_c$ with $\nu_f = 1$ ($T = 1.6$ K). The arrows in **c** denote the critical field $B_{\perp C}$. The arrows in **d** denote the size of the jump in $R_H$ that corresponds to a jump in the Hall density by $n_M$ near $B_{\perp C}$. **e,** Temperature dependence of $R_{xx}$ for $\nu = 1 + 1$ (blue) and $\nu = 0 + 1$ (green) at $B_\perp = 0$ T, as well as the temperature dependence of $R_{bulk}$ for $\nu = 1 + 1$ (orange) at $B_\perp = 0$ T. The dashed line marks $\frac{h}{2e^2}$. **f,** In-plane (black and grey) and out-of-plane (blue) magnetoresistance $R_{xx}(0\,T)/R_{xx}(B)$ at $\nu = 1 + 1$ (black and blue) and $\nu = 1 + 0.8$ (grey) ($T = 1.6$ K).

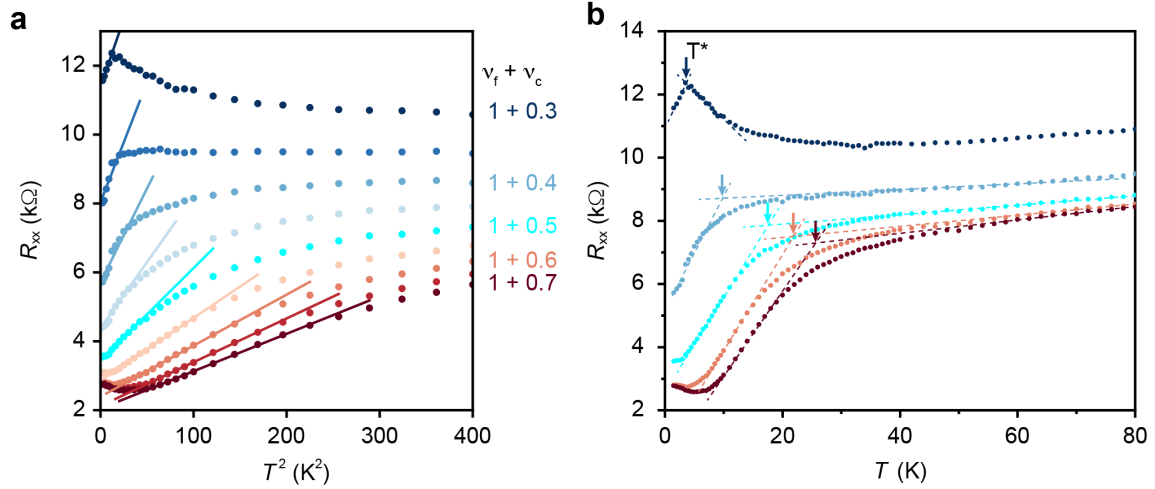

**Extended Data Figure 2 | Determination of $A$ and $T^*$. a,** $R_{xx}$ versus $T^2$ at varying $\nu_c$ in the Kondo lattice region ($B_\perp = 0$ T). The solid lines show the linear fits to the low-temperature part of the data ($R_{xx} = A \times T^2 + R_0$), from which $A$ can be extracted. **b,** $R_{xx}$ versus $T$ at varying $\nu_c$ showing the determination of $T^*$. (Results are from device 1.)

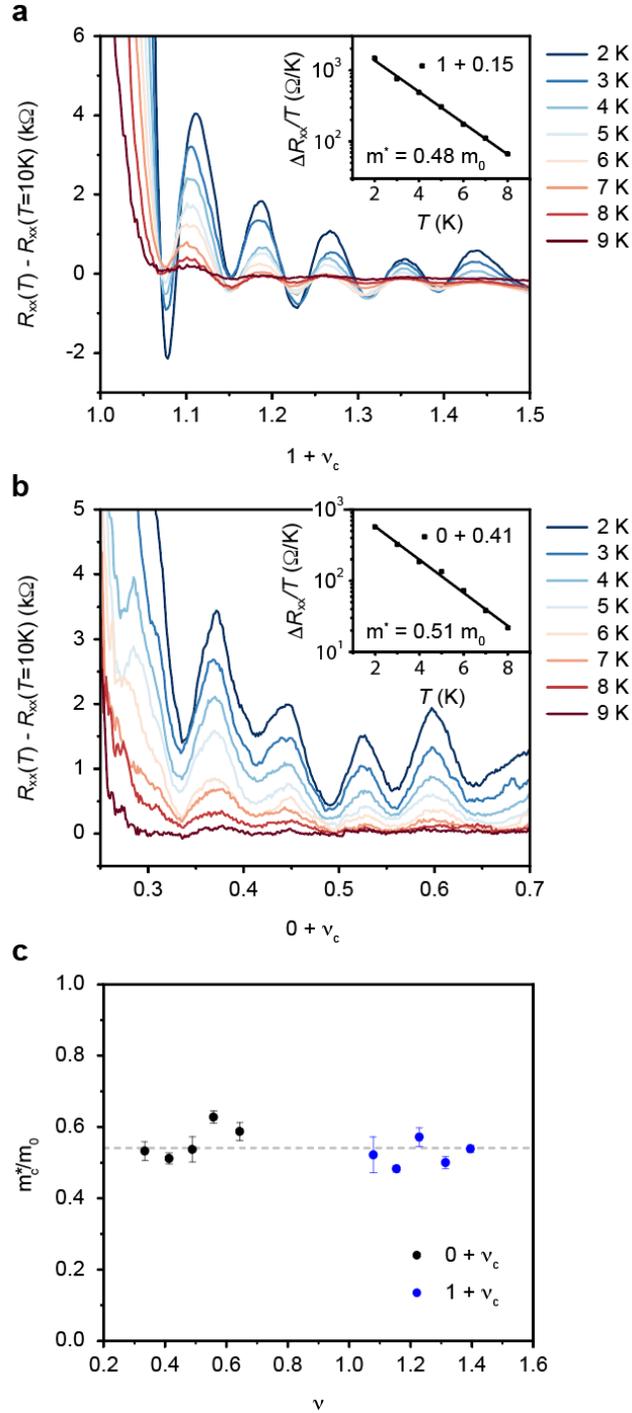

**Extended Data Figure 3 | Determination of the quasiparticle effective mass in the W-layer after Kondo breakdown. a,b,** Filling factor dependence of $R_{xx}(T) - R_{xx}(T = 10\,K)$ at $B_\perp = 14$ T and varying temperatures for $\nu = 1 + \nu_c$ (**a**) and $\nu = 0 + \nu_c$ (**b**). The insets show the corresponding Dingle plots of the SdH amplitude (divided by $T$) as a function of $T$ at selected fillings. The slope of the linear fit (solid line) gives the quasiparticle effective mass $m_c^*$. **c,** Filling factor dependence of the extracted quasiparticle effective mass (in units of the free electron mass $m_0$) for both $\nu = 1 + \nu_c$ (blue) and $\nu = 0 + \nu_c$ (black). (Results are from device 1.)

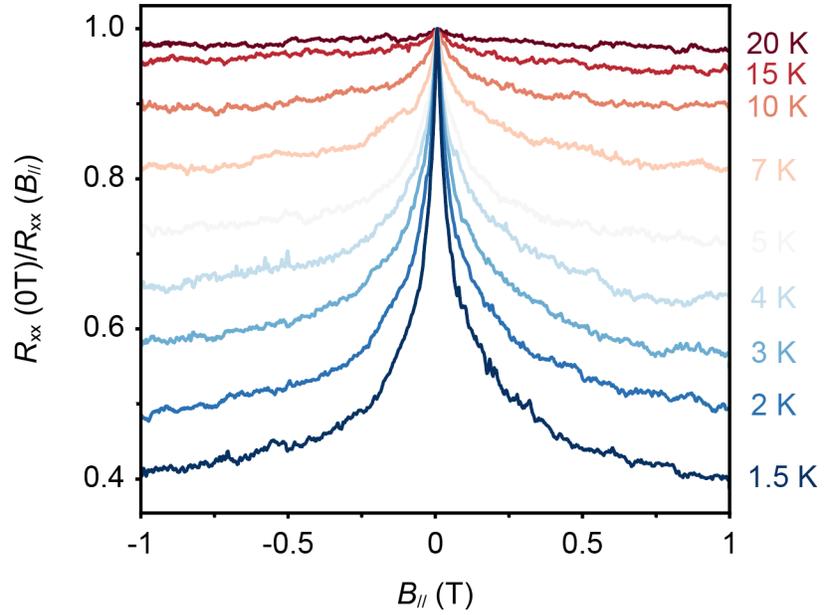

**Extended Data Figure 4 | Temperature dependence of the in-plane magnetoresistance at $\nu = 1+1$.** The in-plane magnetoresistance $R_{xx}(0\,T)/R_{xx}(B_\parallel)$ at varying temperatures. Suppression in $R_{xx}(0\,T)/R_{xx}(B_\parallel)$ is observed up to about $T = 20$ K.

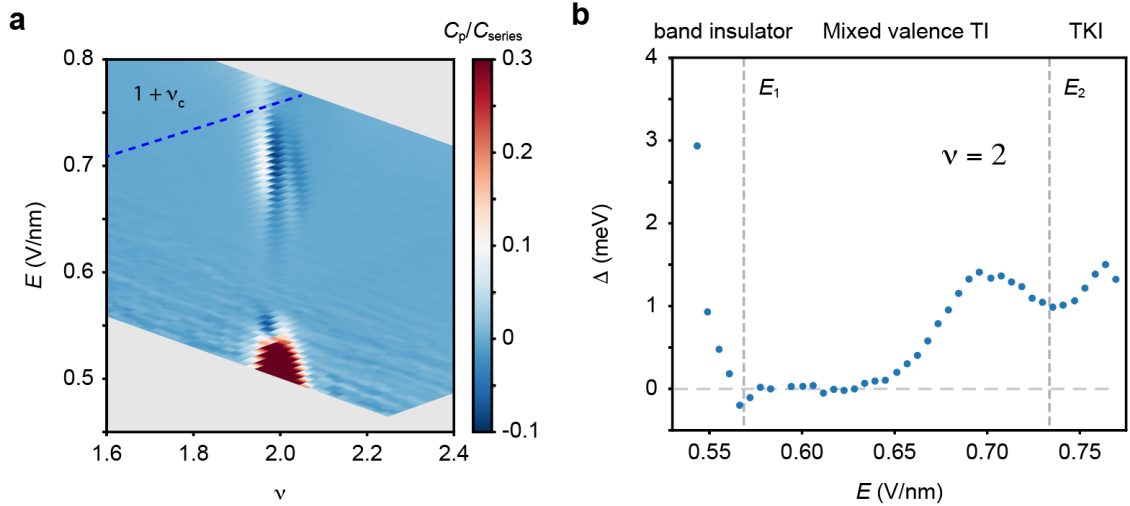

**Extended Data Figure 5 | Penetration capacitance and the gap size at $\nu = 2$. a,** The normalized penetration capacitance $C_p/C_{series}$ as a function of $\nu$ and $E$ at $T = 1.6$ K and $B_\perp = 0$ T. The Kondo lattice region is marked by blue dashed lines. An incompressible state is observed at $\nu = 2$. **b,** Electric field dependence of the thermodynamic gap size $\Delta$ at $\nu = 2$, obtained by integrating the normalized penetration capacitance $C_p/C_{series}$ (see Methods). The grey dashed lines indicate the phase boundaries ($E_1$ and $E_2$, see main text) separating three regimes: band insulator, mixed-valence TI and TKI.

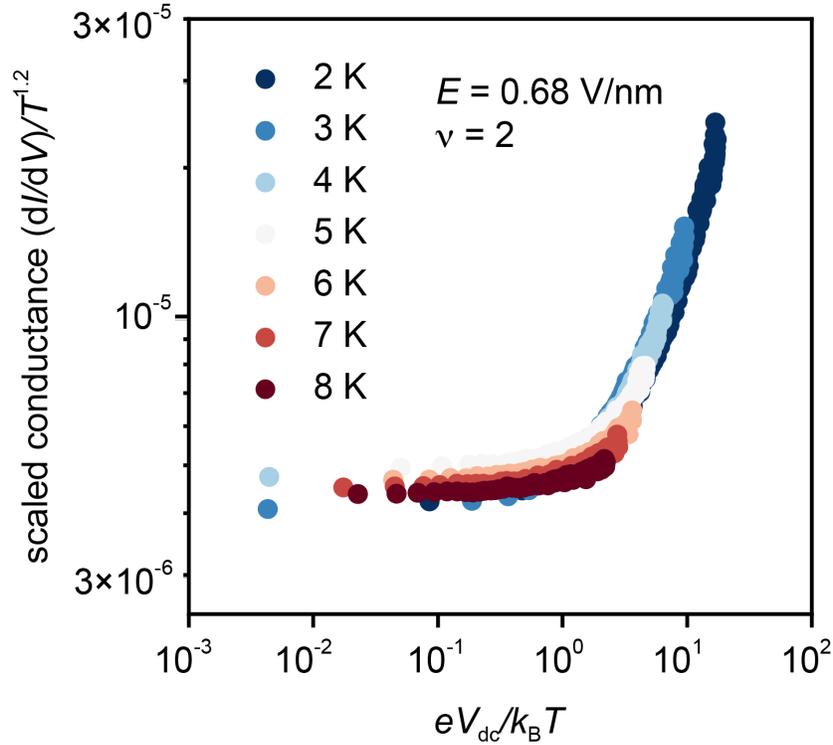

**Extended Data Figure 6 | Evidence of Luttinger-liquid behavior for the mixed-valence TI.** The scaled differential conductance $G_{xx} \equiv (dI/dV)/T^{1.2}$ as a function of $eV_{dc}/k_B T$ at varying temperatures ($V_{dc}$ is the measured dc value of $V_{xx}$ in Fig. 1e). All data collapse to a single curve, consistent with Tomonaga-Luttinger liquid theory, which describes the universal behavior of one-dimensional interacting fermionic system. The experimental hallmark of a Luttinger liquid is the power-law dependence of the conductance on temperature and bias voltage in the limits of $eV_{dc} \ll k_B T$ and $eV_{dc} \gg k_B T$, respectively.